\documentclass{LMCS}

\def\Doi{9(2:02)2013}
\lmcsheading%
{\doi}
{1--17}
{}
{}
{Apr.~\phantom02, 2012}
{Apr.~\phantom02, 2013}
{}

 \usepackage{hyperref,enumerate}
 \usepackage{breakurl}             
\usepackage{amssymb}

\title[The degree structure of Weihrauch reducibility]{The degree structure of Weihrauch reducibility}
\author[K.~Higuchi]{Kojiro Higuchi\rsuper a}
\address{{\lsuper a}Mathematical Institute, Tohoku University,
             Sendai, Miyagi, Japan
}
\email{sa7m24@math.tohoku.ac.jp}
\author[A.~Pauly]{Arno Pauly\rsuper b}
\address{{\lsuper b}Computer Laboratory, University of Cambridge, United Kingdom}
\email{Arno.Pauly@cl.cam.ac.uk}

\keywords{Weihrauch degrees, Medvedev degrees, 
Heyting algebra, superintuitionistic logic, computable analysis}
\ACMCCS{[{\bf Theory of computation}]:  Logic--Modal and temporal
  logics; Logic---Constructive mathematics} 
\subjclass{F.4.1}

\begin{document}
\theoremstyle{definition}
\newtheorem{theorem}{Theorem}
\newtheorem{definition}[theorem]{Definition}
\newtheorem{lemma}[theorem]{Lemma}
\newtheorem{problem}[theorem]{Problem}
\newtheorem{proposition}[theorem]{Proposition}
\newtheorem{corollary}[theorem]{Corollary}
\newtheorem{example}[theorem]{Example}
\newtheorem{observation}[theorem]{Observation}
\newtheorem{deflemma}[theorem]{Definition \& Lemma}
\newcommand{\dom}{\operatorname{dom}}
\newcommand{\Lev}{\operatorname{Lev}}
\newcommand{\range}{\operatorname{range}}
\newcommand{\id}{\textnormal{id}}
\newcommand{\Baire}{{\mathbb{N}^\mathbb{N}}}
\newcommand{\Cantor}{{\{0, 1\}^\mathbb{N}}}
\newcommand{\mto}{\rightrightarrows}
\newcommand{\Sep}{\textnormal{Sep}}
\newcommand{\theo}{\textsc{Th}(\mathfrak{W})}
\newcommand{\llpo}{\textrm{LLPO}}

\begin{abstract}
We answer a question by {\sc Vasco Brattka} and {\sc Guido Gherardi}
by proving that the Weihrauch lattice is not a Brouwer algebra. The
computable Weihrauch lattice is also not a Heyting algebra, but the
continuous Weihrauch lattice is. We further investigate embeddings of the
Medvedev degrees into the Weihrauch degrees.
\end{abstract}

\maketitle

\section{Introduction}
In \cite{gherardi, brattka3} Weihrauch reducibility was suggested as conceptual tool to investigate the computational content of mathematical theorems. A theorem $T$  of the form $\forall x \in X \ \exists y \in Y \ P(x, y)$ (with some arbitrary binary predicate $P$) can be considered as the definition of a multi-valued function $f_T: X \mto Y$ via Skolemization; and the computability of $f_T$ amounts to a form of constructive truth of $T$. If $f_T$ is Weihrauch reducible to $f_S$ derived from some other theorem $S$, then $S$ implies $T$ in a strong constructive sense, as a single invocation of $S$ in an otherwise constructive proof is sufficient to prove $T$. Hence, the degree of incomputability of $f_T$, i.e. its Weihrauch degree, tells us something about how far away from being constructively true the theorem $T$ is -- for some notion of constructive truth. Contributions to this research programme can be found e.g.~in \cite{paulyincomputabilitynashequilibria, paulybrattka3cie, gherardi4, hoyrup2b}.

Another approach to constructive truth are (super)intuitionistic logics, so we would like to know whether these are compatible, i.e. whether we can consider Weihrauch degrees to be the truth-values of a superintuitionistic logic. We point out that Medvedev's original definition of the reducibility named after him was motivated by the desire to identify intuitionistic truth-values with Medvedev degrees \cite{medvedev, terwijn}. The degrees of a reducibility structure can be conceived of as truth-values of a superintuitionistic logic (with the \emph{easy} degrees being closer to truth), if and only if they form a Brouwer algebra \cite{chagrov}. Below we shall demonstrate that neither the whole Weihrauch lattice nor several of its usual modifications form a Brouwer algebra, however, three cases remain open.

Surprisingly, the {\bf dual} of the continuous (i.e. relativized) Weihrauch lattice does turn out to be a Brouwer algebra; this is equivalent to the continuous Weihrauch lattice being a Heyting algebra. In this regard, the continuous Weihrauch lattice exhibits exactly the opposite behaviour of the Medvedev lattice, which is Brouwerian but not Heyting \cite{medvedev, sorbi}. None of the computable versions of the Weihrauch lattice we consider is a Heyting algebra, though.

Apart from the connection to superintuitionistic logics, the presented work also adds to the understanding of the general structural properties of the Weihrauch lattice. As the Weihrauch lattice is not a Brouwer algebra, it cannot be isomorphic to any structure that is. In particular, this yields the result that the Weihrauch lattice is not isomorphic to the Medvedev lattice. In a related fashion, the question whether certain reducibilities induce Brouwer or Heyting algebras have been studied in the literature, e.g. in \cite{kojiro2, shore, simpson4, terwijn2}.

We further study the connection between the Weihrauch lattice and the Medvedev lattice by investigating the properties of certain embeddings between them, both order preserving and order reversing.

\section{Preliminaries}
A partially ordered set $(\mathfrak{L}, \leq)$ is a \emph{lattice}, if one can define operations $\wedge, \vee : \mathcal{L} \times \mathcal{L} \to \mathcal{L}$ (for which we shall use an infix notation) such that for all $\mathbf{a}, \mathbf{b}, \mathbf{c} \in \mathfrak{L}$:
\begin{enumerate}[(1)]
\item $(\mathbf{a} \wedge \mathbf{b}) \leq \mathbf{a}$, $(\mathbf{a} \wedge \mathbf{b}) \leq \mathbf{b}$
\item $\mathbf{c} \leq \mathbf{a}$ and $\mathbf{c} \leq \mathbf{b}$ implies $\mathbf{c} \leq (\mathbf{a} \wedge \mathbf{b})$
\item $\mathbf{a} \leq (\mathbf{a} \vee \mathbf{b})$, $\mathbf{b} \leq (\mathbf{a} \vee \mathbf{b})$
\item $\mathbf{a} \leq \mathbf{c}$ and $\mathbf{b} \leq \mathbf{c}$ implies $(\mathbf{a} \vee \mathbf{b}) \leq \mathbf{c}$
\end{enumerate}
If these operations can be defined, they are determined completely by the partial order, and in turn allow to completely define the partial order themselves. Hence we will either specify the order $\leq$, or the operations $\wedge$, $\vee$, or even neither, if they are clear from the context in the following.

A partially ordered set is \emph{bounded}, if it has a minimal and a maximal element. A lattice is \emph{distributive}, if $\wedge$ and $\vee$ distribute over each other. We call $\mathbf{a} \in \mathfrak{L}$ meet-irreducible, if $\mathbf{a} = \mathbf{b} \wedge \mathbf{c}$ implies $\mathbf{a} = \mathbf{b}$ or $\mathbf{a} = \mathbf{c}$; and dually call $\mathbf{a}$ join-irreducible, if $\mathbf{a} = \mathbf{b} \vee \mathbf{c}$ implies $\mathbf{a} = \mathbf{b}$ or $\mathbf{a} = \mathbf{c}$

A (bounded) lattice $(\mathfrak{L}, \wedge, \vee)$ is a Brouwer algebra, if for all $\mathbf{a}, \mathbf{b} \in \mathfrak{L}$ the set $\{\mathbf{c} \in \mathfrak{L} \mid \mathbf{b} \leq \mathbf{c} \vee \mathbf{a}\}$ contains a smallest element. It is a Heyting algebra, if $\{\mathbf{c} \in \mathfrak{L} \mid \mathbf{c} \wedge \mathbf{a} \leq \mathbf{b}\}$ contains a largest element, this maximal element will be denoted by $\mathbf{a} \rightarrow \mathbf{b}$. Any Brouwer or Heyting algebra is distributive.

For any partially ordered set $(\mathfrak{L}, \leq)$, we use $(\mathfrak{L}, \leq)^{op}$ to denote the partially ordered set $(\mathfrak{L}, \geq)$ where the order is reversed. If $(\mathfrak{L}, \wedge, \vee)$ is a lattice, then so is $(\mathfrak{L}^{op}, \vee, \wedge)$. $\mathfrak{L}$ is distributive and/or bounded, if $\mathfrak{L}^{op}$ is; and $\mathfrak{L}$ is a Brouwer algebra, if and only if $\mathfrak{L}^{op}$ is a Heyting algebra.

We remind the reader that an operation $C : \mathfrak{L} \to \mathfrak{L}$ on a partially ordered set is called a \emph{closure operator}, if $\mathbf{a} \leq C(\mathbf{a})$ holds, if $\mathbf{a} \leq \mathbf{b}$ implies $C(\mathbf{a}) \leq C(\mathbf{b})$ and if $C(C(\mathbf{a})) = C(\mathbf{a})$ holds for all $\mathbf{a}, \mathbf{b} \in \mathfrak{S}$. The elements of the form $C(\mathbf{a})$ are called \emph{fixed} (by $C$).

If $(\mathfrak{L}, \wedge, \vee)$ is a lattice, then for any closure operator $C$ on $\mathfrak{L}$, there is some $\vee'$ defined on $C(\mathfrak{L})$, such that $(C(\mathfrak{L}), \wedge, \vee')$ is a again a lattice. A special case of this are the upper cones, which are of the form $\{\mathbf{a} \in \mathfrak{L} \mid \mathbf{b} \leq \mathbf{a}\}$ for some fixed $\mathbf{b}$. These are the images of the closure operator given as $\mathbf{a} \mapsto \mathbf{a} \vee \mathbf{b}$.

We can interpret every propositional formula as an expression in a Heyting algebra $(\mathfrak{L}, \wedge, \vee, \rightarrow)$ with smallest element $\bot$ and largest element $\top$, considering $\neg \mathbf{a}$ as an abbreviation of $\mathbf{a} \rightarrow \bot$. By $\textsc{Th}(\mathfrak{L})$ we denote the theory of $\mathfrak{L}$, that is the set of all propositional formulae that evaluate to $\top$ regardless of the elements of $\mathfrak{L}$ substituted for the variables. As shown in \cite{chagrov}, $\textsc{Th}(\mathfrak{L})$ is a superintuitionistic logic for any Heyting algebra $\mathfrak{L}$, and any superintuitionistic logic arises as the theory of some Heyting algebra.

A map $m : \mathfrak{L}_1 \to \mathfrak{L}_2$ between two lattices $(\mathfrak{L}_1, \wedge_1, \vee_1)$ and $(\mathfrak{L}_2, \wedge_2, \vee_2)$ is called a meet-semilattice homomorphism, if $m(\mathbf{a} \wedge_1 \mathbf{b}) = m(\mathbf{a}) \wedge_2 m(\mathbf{b})$ holds for all $\mathbf{a}, \mathbf{b} \in \mathfrak{L}_1$; and a join-semilattice homomorphism if $m(\mathbf{a} \vee_1 \mathbf{b}) = m(\mathbf{a}) \vee_2 m(\mathbf{b})$ holds for all $\mathbf{a}, \mathbf{b} \in \mathfrak{L}_1$. If $m$ fulfills both conditions, $m$ is a lattice homomorphism.

If $m : \mathfrak{L}_1 \to \mathfrak{L}_2$ is a lattice homomorphism, $\mathfrak{L}_1$ and $\mathfrak{L}_2$ are Heyting algebras with operations $\rightarrow_1$ and $\rightarrow_2$, and $m$ satisfies $m(\mathbf{a} \rightarrow_1 \mathbf{b}) = m(\mathbf{a}) \rightarrow_2 m(\mathbf{b})$ for all $\mathbf{a}, \mathbf{b} \in \mathfrak{L}_1$, and $m$ also preserves the smallest and the largest element, then $m$ is a Heyting homomorphism. If $m : \mathfrak{L}_1^{op} \to \mathfrak{L}_2^{op}$ is a Heyting morphism, then $m : \mathfrak{L}_1 \to \mathfrak{L}_2$ is a Brouwer morphism.

We call an injective homomorphism an embedding, and remark that if there is a Heyting embedding of $\mathfrak{L}_1$ into $\mathfrak{L}_2$, then $\textsc{Th}(\mathfrak{L}_2) \subseteq \textsc{Th}(\mathfrak{L}_1)$.
\section{The lattices}
Compared to the definition of Weihrauch reducibility in \cite{brattka2}, we shall use a restricted version, which can readily be seen to yield the same degree structure, as every multi-valued function between represented spaces is trivially Weihrauch equivalent to its realizer relation.
\begin{defi}
\label{defweihrauchreduction}
For $P, Q : \subseteq \Baire \mto \Baire$, let $P \leq_W Q$ hold, iff there are computable \linebreak $H, K :\subseteq \Baire \to \Baire$, such that for all choice functions $g$ of $Q$ the function $H\langle \id_\Baire, gK\rangle$ is a choice function of $P$.
\end{defi}

Here the notation $P : \subseteq \Baire \mto \Baire$ identifies $P$ as a partial multi-valued function on Baire space. These can be given by their graphs: Any $G \subseteq \Baire \times \Baire$ defines a partial multi-valued function $P_G$ with $\dom(P_G) = \{x \in \Baire \mid \exists y \ (x, y) \in G\}$ and $P_G(x) = \{y \in \Baire \mid (x, y) \in G\}$ for $x \in \dom(P_G)$.

A partial function $f : \subseteq \Baire \to \Baire$ is a choice function of $P$, if $\dom(f) \supseteq \dom(P)$ and $f(x) \in P(x)$ holds for all $x \in \dom(P)$. Finally, $\langle \ \rangle : \Baire \times \Baire \to \Baire$ is a standard pairing function. Details of Type-2 computability theory can be found in \cite{weihrauchd}. A particular aspect we will use repeatedly is the existence of an effective enumeration $(\Phi_n)_{n \in \mathbb{N}}$ of the partial computable functions with maximal domain.

One can readily verify that $\leq_W$ is transitive and reflexive, hence a preorder. The partially ordered set of equivalence classes induced by $\leq_W$, i.e. of Weihrauch degrees, shall be denoted by $\mathfrak{W}$. Another reducibility we will occasionally refer to is Medvedev reducibility \cite{medvedev}, which is defined for subsets of Baire space via $A \leq_M B$ for $A, B \subseteq \Baire$, if there is a computable function $H$ with $B \subseteq \dom(H)$ and $H(B) \subseteq A$. The Medvedev degrees will be denoted by $\mathfrak{M}$. For details, see \cite{sorbi}.

There are several interesting operations on Weihrauch degrees introduced in \cite{brattka2, paulyreducibilitylattice}. For the proofs that these operations are invariant under Weihrauch reducibility, we refer to these papers.
\begin{defi}
For $P, Q : \subseteq \Baire \mto \Baire$, define $P \coprod Q, P \oplus Q, P \times Q : \subseteq \Baire \mto \Baire$ via $(P \coprod Q)(0p) = 0P(p)$, $(P \coprod Q)(1p) = 1Q(p)$, $(P \oplus Q)(\langle p, q\rangle) = 0P(p) \cup 1Q(q)$ and \linebreak  $P \times Q(\langle p, q\rangle) = \langle P(p), Q(q)\rangle$.
\end{defi}

The multi-valued function with the empty domain $\bot$ is the bottom element. The nature of the top element is more complicated: Based on Definition \ref{defweihrauchreduction} a multivalued function is maximal regarding $\leq_W$, iff it has no choice function. To avoid unwelcome complications, we assume that all multivalued functions on Baire space have choice functions, and obtain a top element by adjoining an artificial degree $\top$. Consequently, we understand $P \coprod \top = \top$, $P \times \top = \top$ and $P \oplus \top = P$. A more detailed discussion on how to treat the top element is relegated to a later publication.

\begin{thm}[{\cite{brattka2, paulyreducibilitylattice}}]
The Weihrauch degrees $\mathfrak{W}$ form a distributive lattice $(\mathfrak{W}, \oplus, \coprod)$.
\end{thm}

The upper cone of the identity $\id_\Baire$ on Baire space is of particular interest for applications, as it contains exactly those multi-valued functions with a computable element in their domain. This is because the computable witness $K$ for $\id_\Baire \leq_W g$ in Definition \ref{defweihrauchreduction} has to produce an element of $\dom(g)$ for each element of $\dom(\id_\Baire)$. Following \cite{paulybrattka}, we call the degrees in this upper cone \emph{pointed}, the corresponding partially ordered set is denoted by $p\mathfrak{W}$. It is easy to see that $\coprod, \oplus, \times$ all preserve pointedness, hence, $(p\mathfrak{W}, \oplus, \coprod)$ is again a distributive lattice.

Two operators on $\mathfrak{W}$ have turned out to be useful in order to characterize concrete problems. Additionally, these operators give rise to further variants of the Weihrauch lattice. The operation $^*$ was introduced in \cite{paulyincomputabilitynashequilibria, paulyreducibilitylattice} as $\overline{\phantom{f}}$, and allows finitely many parallel uses of the initial problem, while $\widehat{\phantom{f}}$ was introduced in \cite{brattka3, brattka2} and allows infinitely many parallel uses of the initial problem. In the following definition, we make use of standard tupling functions $\langle, \rangle$ of both finite and infinite arity.

\begin{defi}
For $P : \subseteq \Baire \mto \Baire$, define $P^*, \widehat{P} : \subseteq \Baire \mto \Baire$ via $P^*(n\langle p_1, \ldots, p_n\rangle) = n\langle P(p_1), \ldots, P(p_n)\rangle$ and $\widehat{P}(\langle p_1, p_2, \ldots \rangle) = \langle P(p_1), P(p_2), \ldots \rangle$. We understand $P^*(0p) = 0^\mathbb{N}$ and $\top^* = \hat{\top} = \top$.
\end{defi}

By \cite[Proposition 4.2]{brattka2} $\widehat{\phantom{f}}$ is a closure operator, and by \cite[Theorem 6.5]{paulyreducibilitylattice} $^*$ is a closure operator. The image of $^*$ allows a nice characterization in terms of $\times$, as we have $\mathbf{f} = \mathbf{f}^*$, if and only if $\mathbf{f} = \mathbf{f} \times \mathbf{f}$ for all $\mathbf{f} \in p\mathfrak{W}$. Note that $0^\mathbb{N} \in \dom(P^*)$ for any $P : \subseteq \Baire \mto \Baire$, hence $\mathfrak{W}^* \subset p\mathfrak{W}$.

The structural role of $\times$ and $^*$ can be described in terms of Kleene algebras, as was observed initially by \textsc{Brattka}. We adapt the definition from \cite{kozen}, and refer to the same work for an overview on the theory of Kleene algebras.
\begin{defi}
A commutative Kleene algebra is a tuple $(\mathfrak{L}, \vee, \cdot, ^*, 0, 1)$, where $(\mathfrak{L}, \vee, 0)$ is a bounded join-semilattice, together with operations $\cdot : \mathfrak{L} \times \mathfrak{L} \to \mathfrak{L}$ and $^* : \mathfrak{L} \to \mathfrak{L}$ and a constant $1 \in \mathfrak{L}$ such that the following axioms hold for all $\mathbf{a}, \mathbf{b}, \mathbf{c} \in \mathfrak{L}$:
\begin{enumerate}[(1)]
\item ($\mathbf{a} \cdot \mathbf{b}) \cdot \mathbf{c} = \mathbf{a} \cdot (\mathbf{b} \cdot \mathbf{c})$
\item $1 \cdot \mathbf{a} = \mathbf{a} \cdot 1 = \mathbf{a}$
\item $\mathbf{a} \cdot (\mathbf{b} \vee \mathbf{c}) = (\mathbf{a} \cdot \mathbf{b}) \vee (\mathbf{a} \cdot \mathbf{c})$
\item $(\mathbf{a} \vee \mathbf{b}) \cdot \mathbf{c}  = (\mathbf{a} \cdot \mathbf{b}) \vee (\mathbf{b} \cdot \mathbf{c})$
\item $0 \cdot \mathbf{a} = \mathbf{a} \cdot 0 = 0$
\item $\mathbf{a} \cdot \mathbf{b} = \mathbf{b} \cdot \mathbf{a}$
\item $1 \vee (\mathbf{a} \cdot \mathbf{a}^*) \leq \mathbf{a}^*$
\item $\mathbf{a} \cdot \mathbf{b} \leq \mathbf{a}$ implies $\mathbf{a} \cdot \mathbf{b}^* \leq \mathbf{a}$
\end{enumerate}
\end{defi}

\begin{lem}
\label{lemmakleenealgebra}
$(\mathfrak{W}, \coprod, \times, ^*, \bot, \id_\Baire)$ is a commutative Kleene algebra.
\end{lem}

\begin{proof}
Properties 1. - 7. follow rather directly from the definitions. For 8., assume that $P \times Q \leq_W P$ is witnessed by computable $H$, $K$. Define $K'$ recursively by $K'\langle p, 0q\rangle = p$ and $K'\langle p, (n+1)\langle q_1, \ldots, q_n, q_{n+1}\rangle\rangle = K(\langle K'\langle p, n\langle q_1, \ldots, q_n\rangle\rangle, q_{n + 1} \rangle)$. Furthermore, define $H'$ recursively by $H'\langle \langle p, 0q\rangle, r\rangle = \langle r, 0^\mathbb{N}\rangle$ and $H'\langle \langle p, (n+1)\langle q_1, \ldots, q_{n+1}\rangle \rangle, r\rangle = \langle h_1, (n+1)\langle h_2^1, \ldots, h_2^{n+1}\rangle\rangle$ where $H\langle \langle K'\langle p, n\langle q_1, \ldots, q_n\rangle\rangle, q_{n+1}\rangle, r\rangle = \langle r', h_2^{n+1}\rangle$ and \linebreak  $H'\langle \langle p, n\langle q_1, \ldots, q_n\rangle \rangle, r''\rangle = \langle h_1, n\langle h_2^1, \ldots, h_2^n\rangle \rangle$. Now $H'$, $K'$ witness $P \times Q^* \leq_W P$.
\end{proof}

The image of a lattice under a closure operator is again a lattice, and a sub-meet-semilattice of the original one, i.e. the binary infima of fixed elements are fixed elements themselves. For both $^*$ and $\widehat{\phantom{f}}$, even the stronger result holds that they commute with $\oplus$ on all degrees. The supremum in the new lattices turns out to be a familiar operation, if we restrict ourselves to pointed degrees: $\mathfrak{W}^*$ and $p\widehat{\mathfrak{W}}$ are lattices with $\oplus$ as infimum and $\times$ as supremum. For $p\widehat{\mathfrak{W}}$ this was proven in \cite{brattka2}. The result for $\mathfrak{W}^*$ will be the consequence of the following observations:

\begin{lem}
\label{lemmaoplusstar}
$(\mathbf{f} \oplus \mathbf{g})^* = \mathbf{f}^* \oplus \mathbf{g}^*$ for all $\mathbf{f}, \mathbf{g} \in \mathfrak{W}$.
\end{lem}

\begin{proof}
For $P, Q : \subseteq \Baire \mto \Baire$ the reduction $(P \oplus Q)^* \leq_W P^*$ is witnessed by computable $H, K : \subseteq \Baire \to \Baire$ defined via $K(n\langle \langle p_1, q_1\rangle, \ldots, \langle p_n, q_n\rangle \rangle) = n\langle p_1, \ldots, p_n\rangle$ and $H(\langle q, n\langle p_1, \ldots, p_n\rangle \rangle) = n\langle 0p_1, \ldots, 0p_n\rangle$. By symmetry, we can conclude $(\mathbf{f} \oplus \mathbf{g})^* \leq_W \mathbf{f}^* \oplus \mathbf{g}^*$.

For the remaining direction, we provide witnesses for $P^* \oplus Q^*
\leq_W (P \oplus Q)^*$. Define computable $K : \subseteq \Baire \to
\Baire$ via 
\[K(\langle n \langle p_1, \ldots, p_n\rangle, m \langle q_1, \ldots,
q_m\rangle\rangle) = (n\cdot m)\langle \langle p_1, q_1\rangle,
\langle p_1, q_2\rangle, \ldots, \langle p_n, q_m\rangle\rangle.
\]
 Further define $H' : \subseteq \Baire \mto \Baire$ via 
\[\eqalign{H'(\langle \langle nq_1, mq_2\rangle, (n \cdot m)\langle
d_{11}p_{11}, d_{12}p_{12}, \ldots, d_{nm}p_{nm}\rangle \rangle)\cr
= \begin{cases} 0n\langle p_{1i_1} \ldots p_{ni_n}\rangle & \forall j
  \leq n \ \exists i_j \ d_{ji_j} = 0 \\ 1m\langle p_{j_11} \ldots
  p_{j_mm} \rangle & \forall i \leq m \ \exists j_i \ d_{j_ii} =
  1 \end{cases}}
\]
 and let $H$ be a computable choice function of $H'$. Then $H$ and $K$
 witness $P^* \oplus Q^* \leq_W (P \oplus Q)^*$.
\end{proof}

\begin{lem}
\label{lemmacoprodstartimes}
$(\mathbf{f} \coprod \mathbf{g})^* = \mathbf{f}^* \times \mathbf{g}^*$ for all $\mathbf{f}, \mathbf{g} \in \mathfrak{W}$.
\end{lem}

\begin{proof}
For $P, Q : \subseteq \Baire \mto \Baire$ the reduction $(P^* \times Q^*) \leq_W (P \coprod Q)^*$ is witnessed by computable $H, K : \subseteq \Baire \to \Baire$ defined via: \[K(\langle n\langle p_1, \ldots, p_n\rangle, m\langle q_1, \ldots, q_m\rangle\rangle) = (n+m)\langle 0p_1, \ldots, 0p_n, 1q_1, \ldots, 1q_m\rangle\] \[H(\langle r, (n+m) \langle 0p_1, \ldots, 0p_n, 1q_1, \ldots, 1q_m\rangle \rangle) = \langle n\langle p_1, \ldots, p_n\rangle, m\langle q_1, \ldots, q_m\rangle \rangle\]
The reduction $(P \coprod Q)^* \leq_W (P^* \times Q^*)$ is witnessed by computable $H, K : \subseteq \Baire \to \Baire$ defined via \[K(n\langle d_1r_1, \ldots, d_nr_n\rangle) = \langle |\{i \mid d_i = 0\}|\langle p_1, \ldots, p_{|\{i \mid d_i = 0\}|}\rangle, |\{i \mid d_i = 1\}|\langle q_1, \ldots, q_{|\{i \mid d_i = 1\}|}\rangle\rangle\] where $p_i = r_{\min \{k \mid i = |\{j \leq k \mid d_j = 0\}|\}}$, $q_i = r_{\min \{k \mid i = |\{j \leq k \mid d_j = 1\}|\}}$; and via \[H(\langle n\langle d_1r_1, \ldots, d_nr_n\rangle, \langle l\langle p_1, \ldots, p_l\rangle, k\langle q_1, \ldots, q_k\rangle\rangle \rangle) = n\langle d_1s_1, \ldots, d_ns_n\rangle\] where $s_i = p_{|\{j \leq i \mid d_j = 0\}|}$ for $d_i = 0$ and $s_i = q_{|\{j \leq i \mid d_j = 1\}|}$ for $d_i = 1$.
\end{proof}

\begin{prop}
$(\mathfrak{W}^*, \oplus, \times)$ is a lattice.
\label{propidempotentlattice}
\end{prop}

\begin{proof}
Lemma \ref{lemmaoplusstar} implies that $\oplus$ is indeed the infimum in $\mathfrak{W}^*$, and Lemma \ref{lemmacoprodstartimes} implies that $\times$ is the supremum in $\mathfrak{W}^*$.
\end{proof}

\begin{cor}
$p\widehat{\mathfrak{W}}$ is a sublattice of $\mathfrak{W}^*$.
\end{cor}

\begin{proof}
This follows from Proposition \ref{propidempotentlattice} together with \cite[Corollary 4.7, Propositions 4.8, 4.9]{brattka2}.
\end{proof}

\begin{prop}
$\mathbf{f} \oplus (\mathbf{g} \times \mathbf{h}) = (\mathbf{f} \oplus \mathbf{g}) \times (\mathbf{f} \oplus \mathbf{h})$ and $\mathbf{f} \times (\mathbf{g} \oplus \mathbf{h}) = (\mathbf{f} \times \mathbf{g}) \oplus (\mathbf{f} \times \mathbf{h})$ for all $\mathbf{f} \in \mathfrak{W}^*$, $\mathbf{g}, \mathbf{h} \in \mathfrak{W}$.
\end{prop}

\begin{proof}
For any $P, Q, R : \subseteq \Baire \mto \Baire$ the reduction \ldots

\begin{enumerate}[(1)]
\item $P \oplus (Q \times R) \leq_W (P \oplus Q) \times (P \oplus R)$ is witnessed by computable $H, K$ defined via $K(\langle p, \langle q, r\rangle\rangle) = \langle \langle p, q\rangle, \langle p, r\rangle\rangle$ and $H(\langle x, \langle 1q, 1r\rangle\rangle) = 1\langle q, r\rangle$ as well as $H(\langle x, \langle d_1p_1, d_2p_2\rangle\rangle) = 0p_i$ with $i$ such that $d_i = 0$.

\item $(P \oplus Q) \times (P \oplus R) \leq_W (P \times P) \oplus (Q \times R)$ is witnessed by computable $H, K$ defined via $K(\langle \langle p_1, q\rangle, \langle p_2, r\rangle\rangle) = \langle \langle p_1, p_2\rangle, \langle q, r\rangle\rangle$ and $H(\langle x, d\langle p, q\rangle\rangle) = \langle dp, dq\rangle$.

\item $P \times (Q \oplus R) \leq_W (P \times Q) \oplus (P \times R)$ is witnessed by computable $H, K$ defined via $K(\langle p, \langle q, r\rangle\rangle) = \langle \langle p, q\rangle, \langle p, r\rangle\rangle$ and $H(\langle x, d\langle p, q\rangle\rangle) = \langle p, dq\rangle$.

\item $(P \times Q) \oplus (P \times R) \leq_W (P \times P) \times (Q \oplus R)$ is witnessed by computable $H, K$ defined via $K(\langle \langle p_1, q\rangle, \langle p_2, r\rangle\rangle) = \langle \langle p_1, p_2\rangle, \langle q, r\rangle\rangle$ and $H(\langle x, \langle \langle p_1, p_2\rangle, dq\rangle\rangle) = d\langle p_{d+1},q\rangle$.
\end{enumerate}

\noindent If $P$ is a representative of some $f \in \mathfrak{W}^*$, then $P \times P \equiv_W P$, so the first claim follows from 1. and 2., and the second from 3. and 4..
\end{proof}

\begin{cor}
Both $\mathfrak{W}^*$ and $p\widehat{\mathfrak{W}}$ are distributive.
\end{cor}

As computable functions are by necessity continuous, and moreover, proofs for Weihrauch reducibility tend to employ a blend of recursion theoretic and topological arguments, a straight-forward generalization of Definition \ref{defweihrauchreduction} is the following:

\begin{defi}
\label{defweihrauchreductioncontinuous}
For $P, Q : \subseteq \Baire \mto \Baire$, let $P \leq_W^c Q$ hold, iff there are continuous \linebreak $H, K :\subseteq \Baire \to \Baire$, such that for all choice functions $g$ of $Q$ the function $H\langle \id_\Baire, gK\rangle$ is a choice function of $P$.
\end{defi}

It is well-known that Definition \ref{defweihrauchreductioncontinuous} can also be obtained from Definition \ref{defweihrauchreduction} via relativization with respect to an arbitrary oracle. All the result above relativize, yielding corresponding statements about the continuous Weihrauch degrees. Note that the relativization of pointedness is non-emptiness of the domain, hence, the pointed continuous Weihrauch degrees are all continuous degrees but $\bot$. We shall use $\mathfrak{C}$ to denote the non-empty continuous Weihrauch degrees, and write $\mathfrak{C}_0$ if $\bot$ is included. We consider the element $\top$ to be present in $\mathfrak{C}$, $\mathfrak{C}_0$, and to be identical to the top element of $\mathfrak{W}$.
\begin{prop}
$(\mathfrak{C}_0, \oplus, \coprod)$, $(\mathfrak{C}, \oplus, \coprod)$, $(\mathfrak{C}^*, \oplus, \times)$ and $(\widehat{\mathfrak{C}}, \oplus, \times)$ are distributive lattices. $\mathfrak{C}$ is a sublattice of $\mathfrak{C}_0$ and $\widehat{\mathfrak{C}}$ is a sublattice of $\mathfrak{C}^*$.
\end{prop}

In the continuous case, the coproduct $\coprod$ can easily be extended to a countable number of arguments, as can be done for the infimum $\oplus$ and the product $\times$. Hence, the lattices in the continuous case are $\aleph_0$-complete. Also, using the countable version of $\coprod$ the relativization of Lemma \ref{lemmakleenealgebra} can be strengthened to yield a closed semiring. These features are not available for the computable reductions, as the following non-relativizing proofs show:

\begin{prop}
\label{propnoinfinitesuprema}
$\mathfrak{W}$ has no non-trivial infinite suprema, i.e.~a sequence $(\mathbf{a}_i)_{i \in \mathbb{N}}$ of degrees in $\mathfrak{W}$ has a supremum if and only if it is already the supremum of some finite subset $(\mathbf{a}_i)_{i \leq N}$.
\end{prop}

\begin{proof}
As we have finite suprema, it is sufficient to show that no countable strictly increasing sequence $(P_e :\subseteq \Baire \mto \Baire)_{e \in \mathbb{N}}$ can admit a supremum. Let $Q$ be an upper bound for the $P_e$. We construct an $R$ satisfying $P_e \leq_W R$ for all $e \in \mathbb{N}$, but $Q \nleq_W R$, by defining $R(a_np) = P_n(p)$ for a sequence $(a_n)_{n \in \mathbb{N}}$ to be determined next. In particular, it is clear from the definition that $P_e \leq_W R$ holds for any $e \in \mathbb{N}$.

Now we define the $a_n$ recursively in stages $n \in \mathbb{N}$, using as auxiliary $a_{-1} = 0$. Let $\Phi_{n}$ be the $n$th partial computable function. If there is a $q \in \dom(Q) \cap \dom(\Phi_n)$ with $\Phi_n(q)(0) > a_{n-1}$, set $a_n = \Phi_n(q)(0) + 1$ for such a $q$; otherwise let $a_n = a_{n-1} + 1$.

Now assume that $Q \leq_W R$ were witnessed by $H$, $\Phi_n$. This directly implies $\dom(Q) \subseteq \dom(\Phi_n)$. So if there were a $q \in \dom(Q)$ with $\Phi_n(q)(0) > a_{n-1}$, we would find $a_n > \Phi_n(q)(0) > a_{n-1}$, hence $\Phi_n(q) \notin \dom(R)$, violating the assumption that $\Phi_n$ witnesses a reduction to $R$. So we may conclude $\Phi_n(q)(0) < a_n$ for all $q \in \dom(Q)$. This in turn means that $H$, $\Phi_n$ witness the reduction $Q \leq_W R'$, where $R'$ is defined via $R'(a_ip) = P_i(p)$ for $i < n$. As the $P_e$ are increasing, we have $Q \leq_W R' \leq_W P_{n-1}$; which contradicts the further assumption that the $P_e$ are strictly increasing and $P_n \leq_W Q$ holds.
\end{proof}

\begin{prop}
\label{proposition:infiniteinfima}
A sequence $(P_n : \subseteq \Baire \mto \Baire)_{n \in \mathbb{N}}$ with $P_{n + 1} <_W P_{n}$ and $\dom(P_{n + 1}) \leq_M \dom(P_n)$ for all $n \in \mathbb{N}$ has no infimum in $\mathfrak{W}$.
\end{prop}

\begin{proof}
Given such a sequence and $Q : \subseteq \Baire \mto \Baire$ such that $Q \leq_W P_n$ for all $n \in \mathbb{N}$,
we can find $R$ such that $R \leq_W P_n$ for all $n \in \mathbb{N}$ and $R \nleq_W Q$.

We may safely assume that $0f\not\in\dom(P_n)$ for all
$f\in\mathbb{N}^\mathbb{N}$ and $n\in\mathbb{N}$.
For natural numbers $n\in\mathbb{N}$ and $m<n$, let $H^n_m,K^n_m$ be computable functions such that $P_n \leq_W P_m$ via $H^n_m,K^n_m$.
Note that $\dom(P_m) \leq_M \dom(P_n)$ via $K^n_m$.
We construct $(R_n)_{n\in\mathbb{N}}$ stage by stage and define
$R=\bigcup_{n\in\mathbb{N}}R_n$.
$R$ is forced to satisfy $\neg(R\leq_W Q\ \mbox{via}\ \Phi_i, \Phi_j)$ by the
construction of $R_{\langle i, j\rangle}$. This involves the construction of a sequence of natural numbers, as auxiliary assumption we shall understand $a_{-1} = 0$.

{\bf Stage $n=\langle i,j\rangle$}. Assume that we have a finite sequence $(a_k)_{k<n}$ of
natural numbers by this stage $n$.
If there is a choice function $q$ of $Q$ such that
\begin{equation}\label{eq1}(\Phi_j\circ\langle {
\id}_{\mathbb{N}^\mathbb{N}},q\circ\Phi_i\rangle)\langle K^n_0(f),\cdots,K^n_{n-1}(f),f,0^\mathbb{N},0^\mathbb{N},0^\mathbb{N},\cdots\rangle (0)>\max_{k<n}
a_k\end{equation} for some $f\in\dom(P_n)$,
take such $f,q$ and define
\begin{equation}\label{eq2} a_n=(\Phi_j\circ\langle {\id}_{\mathbb{N}^\mathbb{N}},q\circ\Phi_i\rangle)\langle K^n_0(f),\cdots,K^n_{n-1}(f),f,0^\mathbb{N},0^\mathbb{N},0^\mathbb{N},\cdots\rangle (0)+1.\end{equation}
Otherwise, define $a_n=\max_{k<n} a_k + 1$.
Define $R_n$ by
\begin{align*}
R_n &
(\langle K^n_0(f),\cdots,K^n_{n-1}(f),f,0^\mathbb{N},0^\mathbb{N},0^\mathbb{N},\cdots \rangle)
\\
& =\bigcup_{k<n}a_kP_k(K^n_k(f))\cup
a_nP_n(f)\cup\bigcup_{m\in\mathbb{N}\setminus\{0\}} \left (\bigcup_{g\in{\rm
dom}(P_{n+m})}(a_n+m)P_{n+m}(g)\right )
\end{align*}
for all $f\in\dom(P_n)$. Note that $R = \bigcup_{n \in \mathbb{N}} R_n \equiv_W \coprod_{n \in \mathbb{N}} R_n$.

First we claim $R \nleq_W Q$. Assume contrarily that $R \leq_W Q$ is witnessed by $\Phi_i$, $\Phi_j$. Then $\Phi_i$, $\Phi_j$ also witness $R_{\langle i, j\rangle} \leq_W Q$. If there were $q$, $f$ s.t. the inequality (\ref{eq1}) were true, then the definition (\ref{eq2}) shows that $\Phi_i$, $\Phi_j$ do not act correctly. Hence, $\Phi_i$, $\Phi_j$ also have to witness $R'_{\langle i, j\rangle} \leq_W Q$ for $R'_n$ defined via $R'_n(\langle K^n_0(f),\cdots,K^n_{n-1}(f),f,0^\mathbb{N},0^\mathbb{N},0^\mathbb{N},\cdots \rangle) = \bigcup_{k<n}a_kP_k(K^n_k(f))$. But by definition of the $K^n_k$, we find $P_n \leq_W R'_n$. Taking everything together, the assumption $R \leq_W Q$ implies $P_n \leq_W Q$ for some $n$, violating the original assumptions.

Now we show that $R \leq_W P_m$ for any $m\in\mathbb{N}$. We have $R_n \leq_W P_{n+m}$ for any $m \in \mathbb{N} \setminus\{0\}$ since $\dom(P_n) \geq_M \dom(P_{n+m})$ - the reduction is not necessarily uniform in $n$, $m$, though!

Fix $m \in \mathbb{N}$. Note that
$$R=\bigcup_{n\in\mathbb{N}}R_n\equiv_WR_0\coprod R_1\coprod\cdots\coprod
R_{m}\coprod(\bigcup_{n > m}R_n).$$
Since $R_n \leq_W P_m$ for any $n<m$ as shown above and $R_m \leq_W P_m$ due to the $a_mP_m(f)$ term in the definition of $R_m$, it suffices to prove that
$$\bigcup_{n> m}R_n\le_W P_m.$$
This follows from
$$a_mP_m(K_m^n(f)) \in \left ( \bigcup_{k > m}R_k \right ) (\langle K^n_0(f),\cdots,K_m^n(f), \cdots, K^n_{n-1}(f),f,0^\mathbb{N},0^\mathbb{N},0^\mathbb{N},\cdots \rangle)$$
holding for $n > m$.
\end{proof}

\begin{cor}
$\mathfrak{W}$ is not an $\aleph_0$-complete meet-semilattice nor an $\aleph_0$-complete join-semi\-lattice.
\end{cor}

\begin{cor}
$p\mathfrak{W}$ has no non-trivial infinite infima, i.e.~a sequence $(\mathbf{a}_i)_{i \in \mathbb{N}}$ of degrees in $\mathfrak{W}$ has an infimum if and only if it is already the infimum of some finite subset $(\mathbf{a}_i)_{i \leq N}$..
\end{cor}

The domains of any decreasing chains with an infimum in $\mathfrak{W}$ must be an increasing chain in $\mathfrak{M}$, and the following example demonstrates that this case actually occurs:

\begin{exa}
There exists a decreasing sequence $\{{\bf a}_n\}_{n\in\mathbb{N}}$ of degrees
which have its g.l.b. ${\bf a}$ in ${\mathfrak W}$.
\end{exa}

\begin{proof}
Take $f_0<_Tf_1<\cdots$.
Define $P_n(g)=0^\mathbb{N}$ for any $n\in\mathbb{N}$ and $g\not\le_Tf_n$ and
define $P(h)=0^\mathbb{N}$ for any $h\in\mathbb{N}^\mathbb{N}$ such that
$h\not\le_Tf_m$ for all $m\in\mathbb{N}$.
Clearly, $P <_W P_{n+1} <_W P_n$.
Suppose that $Q$ satisfies $Q\le_WP_n$ for all $n\in\mathbb{N}$.
Then, ${\rm dom}(Q)\subseteq {\rm dom}(P)$ -- to see this, assume $h \in \dom(Q)$. Since ${\rm dom}(Q)\ge_M{\rm
dom}(P_n)$ for any $n\in\mathbb{N}$, there is a $q_n \in \dom(P_n)$ with $q_n \leq_T h$ and $q_n \nleq_T f_n$, which in turn implies $h \nleq_T f_n$, i.e. $h \in \dom(P)$.
Let $H$ be a computable function such that $Q\le_WP_0$ via $H,K$ for some $K$.
We have $Q\le_WP$ via $H, {\id}_{\mathbb{N}^\mathbb{N}}$.
\end{proof}

\section{Main results}
In this section we investigate whether any of the lattices introduced in the previous section is a Brouwer or a Heyting algebra. First, we fix some notation: We define the computable shift function $\textsc{Sh} : \Baire \to \Baire$ via $\textsc{Sh}(np) = p$ and denote the Turing jump by $J : \Baire \to \Baire$, i.e.~$J(p)$ is the Halting problem for Turing machines with oracle $p$ considered as a sequence in $\{0, 1\}^\mathbb{N} \subseteq \Baire$.

\begin{thm}
\label{notbrouwer1}
Neither $\mathfrak{W}$ nor $p\mathfrak{W}$ is a Brouwer algebra.
\end{thm}

\begin{proof}
Let $(q_i)_{i \in \mathbb{N} \setminus \{0\}}$ be a sequence of pairwise Turing incomparable elements of Baire space below $J(0^\mathbb{N})$. Define $p_i = i0q_i$, if $iq_i \in \dom(\Phi_i)$ and $\Phi_i(iq_i)(2) > 0$ and $p_i = i1q_i$ otherwise. Now define $\overline{P}, \overline{Q} : \subseteq \Baire \to \Baire$ via $\overline{P}(0^\mathbb{N}) = 0^\mathbb{N} = \overline{Q}(0^\mathbb{N})$ and $\overline{P}(p) = J(0^\mathbb{N}) = \overline{Q}(q)$ for $p \in \{p_i \mid i \in \mathbb{N}\}$, $q \in \{iq_i \mid i \in \mathbb{N}\}$.

Now the set $\{R \in \mathfrak{W} \mid \overline{Q} \leq_W \overline{P} \coprod R\}$ has no minimal element, and no minimal pointed element. Assume the contrary, let $\overline{R}$ be a minimal (pointed) element, and let $\overline{Q} \leq_W \overline{P} \coprod \overline{R}$ be witnessed by computable $H'$, $K' = \Phi_e$.

\begin{enumerate}[(1)]
\item {\bf Claim:} $\Phi_e(eq_e)$ is defined, and $\textsc{Sh}(\Phi_e(eq_e)) \in \dom(\overline{R})$.

As we have $eq_e \in \dom(\overline{Q})$, we find $\Phi_e(eq_e) \in \dom(\overline{P} \coprod \overline{R})$, in particular $\Phi_e(eq_e)(0) \in \{0, 1\}$. Assume $\Phi_e(eq_e)(0) = 0$. Then $\textsc{Sh}(\Phi_e(eq_e)) \in \dom(\overline{P})$, so either $\Phi_e(eq_e) = 0^\mathbb{N}$ or $\Phi_e(eq_e) = 0p_i$ for some $i \in \mathbb{N}$. If $\Phi_e(eq_e) = 0^\mathbb{N}$, then $(\overline{P} \coprod \overline{R})(\Phi_e(eq_e)) = 0^\mathbb{N}$, a contradiction to the assumption that $J(0^\mathbb{N}) = \overline{Q}(eq_e)$ is Turing below $\langle eq_e,(\overline{P} \coprod \overline{R})(\Phi_e(eq_e))\rangle$.

If $\Phi_e(eq_e) = 0p_i$ for $i \neq e$, the Turing incomparability of $q_e$ and $q_i$ is contradicted. So we are left with the case $\Phi_e(eq_e) = 0p_e$. But this contradicts the definition of $p_e$. So the assumption $\Phi_e(eq_e)(0) = 0$ was wrong, which implies $\Phi_e(eq_e)(0) = 1$. Then $\Phi_e(eq_e) \in \dom(\overline{P} \coprod \overline{R})$ yields the claim.

\item  {\bf Claim:} For all $p \in \overline{R}(\textsc{Sh}(\Phi_e(eq_e)))$ we have $J(0^\mathbb{N}) \leq_T \langle q_e, p\rangle$.

\item  {\bf Claim:} Define $\overline{S}$ via $\overline{S}(p) = \overline{Q}(p)$ for all $p \in \dom(Q) \setminus \{eq_e\}$. Then $\overline{R} \nleq_W \overline{S}$.

Assume $\overline{R} \leq_W \overline{S}$ witnessed by computable $G, L$. Then, by $1.$, $L(\textsc{Sh}(\Phi_e(eq_e))) \in \dom(\overline{S})$. By assumption of Turing incomparability, we cannot have $L(\textsc{Sh}(\Phi_e(eq_e))) = iq_i$ for $i \neq e$. Thus, $L(\textsc{Sh}(\Phi_e(eq_e))) = 0^\mathbb{N}$ follows. Hence, the assumption leads us to $G(\langle \textsc{Sh}(\Phi_e(eq_e)), 0^\mathbb{N}\rangle) \in \overline{R}(\textsc{Sh}(\Phi_e(eq_e)))$. Together with $2.$ this contradicts $J(0^\mathbb{N}) \nleq_T q_e$.

\item  {\bf Claim:} $\overline{Q} \leq_W \overline{P} \coprod \overline{S}$.

We give computable witnesses $H, K$ for the claim. Define $H(\langle p, dq\rangle) = q$, $K(lp) = 1lp$ for $l \neq e$ and $K(ep) = 0ejp$ where $j \in \{0, 1\}$ satisfies $p_e = ejq_e$. Intuitively, the reduction calls $\overline{S}$ for any input to $\overline{Q}$ which is also in the domain of $\overline{S}$, where $\overline{S}$ and $\overline{Q}$ agree. The only remaining input is $eq_e$, in which case $\overline{P}$ is called on input $p_e$, which correctly produces $J(0^\mathbb{N})$.
\end{enumerate}

\noindent The assumption $\overline{R}$ were the minimal (pointed) element in $\{R \in \mathfrak{W} \mid \overline{Q} \leq_W \overline{P} \coprod R\}$ is refuted by the construction of the pointed element $\overline{S}$ not above it.
\end{proof}

A similar argument can be used for $\mathfrak{W}^*$ and $p\widehat{\mathfrak{W}}$:

\begin{thm}
\label{notbrouwer2}
Neither $\mathfrak{W}^*$ nor $p\widehat{\mathfrak{W}}$ is a Brouwer algebra.
\end{thm}

\begin{proof}
As in the proof of Theorem \ref{notbrouwer1}, let $(q_i)_{i \in \mathbb{N} \setminus \{0\}}$ be a sequence of pairwise Turing incomparable elements of Baire space below $J(0^\mathbb{N})$. For each $e\in\mathbb{N}\setminus\{0\}$, define a natural number $a_e$ such that
$a_e = p_i(0) + 1$, if $\Phi_e(\langle eq_e,eq_e,\cdots\rangle)=\langle\langle 0^\mathbb{N},0^\mathbb{N},\cdots,0^\mathbb{N},p_i,p_{i+1},\cdots \rangle, r \rangle$ and $p_i\neq 0^\mathbb{N}$, and $a_e = 1$ otherwise.
Define $P,Q:\subseteq\mathbb{N}^\mathbb{N}\to\mathbb{N}^\mathbb{N}$
via $P(0^\mathbb{N})=0^\mathbb{N}=Q(0^\mathbb{N})$ and
$P(p)=J(0^\mathbb{N})=Q(q)$ for $p\in\{a_iq_i\mid
i\in\mathbb{N}\setminus\{0\}\}$, $q\in\{iq_i\mid
i\in\mathbb{N}\setminus\{0\}\}$.

We show that the set $\{R\in\mathfrak{W}\mid\widehat{Q}\leq_W\widehat{P}\times R\}$ does not have a minimal element, and also does not have a minimal parallelizable element.
Take any $\overline{R}$ and $H^\prime,K^\prime=\Phi_e$ such that $\widehat{Q}\leq_W\widehat{P}\times\overline{R}$ via $H^\prime,K^\prime$.
\begin{enumerate}[(1)]
\item {\bf Claim:}  There is $r\in\mathbb{N}^\mathbb{N}$ such that $\Phi_e(\langle eq_e,eq_e,\cdots\rangle)=\langle\langle 0^\mathbb{N},0^\mathbb{N},\cdots \rangle, r \rangle$.\\
Since $\langle eq_e,eq_e,\cdots\rangle\in \dom(\widehat{Q})$, we have $\Phi_e(\langle eq_e,eq_e,\cdots\rangle) \in \dom(\widehat{P}\times\overline{R})$.
Assume $\Phi_e(\langle eq_e,eq_e,\cdots\rangle)=\langle\langle 0^\mathbb{N},0^\mathbb{N},\cdots,0^\mathbb{N},p_i,p_{i+1},\cdots \rangle, r \rangle$ for some $p_i\in\dom(P)\setminus\{0^\mathbb{N}\}$.
Then $p_i = a_mq_m$ for some $m \neq 0$ and $q_m \leq_T q_e$. Since $a_e > p_i(0) = a_m$ holds by definition, we conclude $m \neq e$. But by our choice of the sequence $(q_i)_{i \in \mathbb{N} \setminus \{0\}}$, we have $q_m |_T q_e$ for $m \neq e$, whence we obtain  a contradiction.
\item  {\bf Claim:} For all $f\in\overline{R}(r)$, $J(0^\mathbb{N})\leq_T f$.
\item  {\bf Claim:} Define $S$ via $S(s)=Q(s)$ for all $s\in\dom(Q)\setminus\{eq_e\}$. Then $\overline{R}\not\leq_W\widehat{S}$.

Assume $\overline{R}\leq_W\widehat{S}$ witnessed by computable $G,L$.
Then, by 1, $L(r)\in\dom(\widehat{S})$ and, therefore, $L(r)=\langle 0^\mathbb{N}, 0^\mathbb{N},\cdots\rangle$ by assumption of Turing incomparability.
Hence, we have $G(\langle r, \langle 0^\mathbb{N},0^\mathbb{N},\cdots \rangle \rangle)\in\overline{R}(r)$.
Together with 2, this contradicts $J(0^\mathbb{N})\not\leq_Tq_e$.
\item  {\bf Claim:} $\widehat{Q}\leq_W\widehat{P}\times\widehat{S}$.\\
We give computable witnesses $H,K$ for the claim.
Define $K$ by
$$K(\langle k_1\alpha_1,k_2\alpha_2,\dots\rangle)=\langle\langle\beta_1,\beta_2,\cdots\rangle, \langle\gamma_1,\gamma_2,\cdots\rangle\rangle,$$
where $\beta_i=a_e\alpha_i\ \&\ \gamma_i=0^\mathbb{N}$ if $k_i=e$ and $\beta_i=0^\mathbb{N}\ \&\ \gamma_i=k_i\alpha_i$ if $k_i\neq e$ and define $H$ by
$$H(\langle k_1\alpha_1,k_2\alpha_2,\dots\rangle, \langle\langle\delta^1_1,\delta^1_2,\cdots\rangle, \langle\delta^2_1,\delta^2_2,\cdots\rangle\rangle)=\langle\delta^{l_1}_1,\delta^{l_2}_2,\cdots\rangle,$$
where $l_i=1$ if $k_i=e$ and $l_i=2$ if $k_i\neq e$.
\end{enumerate}
We know that $\overline{R}$ is neither the minimal element nor the minimal parallelizable element of $\{R\in\mathfrak{W}\mid\widehat{Q}\le_W\widehat{P}\times R\}$ by 3 and 4.
This concludes the proof for $\widehat{\mathfrak{W}}$. Regarding $p\mathfrak{W}^*$, note that $\overline{P}$, $\overline{Q}$, $S$ and $\widehat{S}$ constructed above are all pointed. In particular, $\widehat{S}$ is fixed by $^*$, showing that no least element fixed by $^*$ can exist in $\{R\in\mathfrak{W}\mid\widehat{Q}\le_W\widehat{P}\times R\}$.
\end{proof}

The recursion-theoretic methods employed in the proofs of the Theorems \ref{notbrouwer1}, \ref{notbrouwer2} cannot be extended to the continuous case: All elements of Baire space are equivalent with respect to some oracle, hence all the functions used there are equivalent with respect to continuous Weihrauch reducibility. In the following we shall use combinatorial principles to derive an alternative proof technique. The study of continuous Weihrauch reductions between multi-valued functions of this kind was initiated in \cite{weihrauchc}, and extended significantly in \cite{mylatzb}.

\begin{defi}[{\cite{weihrauchc}}]
For $n > 1$, define $\llpo_{n,1} : \subseteq \Baire \mto \Baire$ via $$\dom(\llpo_{n,1}) = \{\langle p_1, p_2, \ldots\rangle \mid 1 \geq |\{\langle i, j\rangle \mid p_i(j) \neq 0\}| \}$$ and $i0^\mathbb{N} \in \llpo_{n, 1}(\langle p_1, p_2, \ldots\rangle)$, if $i \leq n$ and $p_i = 0^\mathbb{N}$.
\end{defi}

\begin{defi}[{\cite{mylatzb}}]
For $n \in \mathbb{N}$, define $\llpo_{\infty, n} : \subseteq \Baire \mto \Baire$ via $$\dom(\llpo_{\infty,n}) = \{\langle p_1, p_2, \ldots\rangle \mid n \geq |\{\langle i, j\rangle \mid  p_i(j) \neq 0\}| \}$$ and $i0^\mathbb{N} \in \llpo_{\infty, n}(\langle p_1, p_2, \ldots\rangle)$, if $p_i = 0^\mathbb{N}$.
\end{defi}

\begin{defi}
For $k \in \mathbb{N}$, define $\sum_{k}^\infty \llpo : \subseteq \Baire \mto \Baire$ via $\sum_{k}^\infty \llpo(\langle 0^\mathbb{N}, p\rangle) = \llpo_{\infty, 2}(p)$, $\sum_{k}^\infty \llpo(\langle 0^n10^\mathbb{N}, p\rangle) = \llpo_{n, 1}(p)$ for $n \geq k$.
\end{defi}

For $n > 1$, by \cite[Satz 4.3]{weihrauchc} we know $\llpo_{n+1, 1} <_W^c \llpo_{n, 1}$, and by \cite[Satz 18]{mylatzb} we know $\llpo_{n, 1} \nleq_W^c \llpo_{\infty, k}$ and $\llpo_{\infty, k + 1} \nleq_W^c \llpo_{n, 1}$ for any $k \in \mathbb{N}$. Additionally, we will need the following:

\begin{lem}
\label{lemmallpo}
$\llpo_{k, 1} \nleq_W^c \sum_{k+1}^\infty \llpo$.
\end{lem}

\begin{proof}
Assume the contrary, witnessed by continuous functions $H$, $K$ with \linebreak $K(p) = \langle K_1(p), \langle K_2^1(p), K_2^2(p), \ldots\rangle\rangle$.
\begin{enumerate}[(1)]
\item For all $n > 1$, $\langle \langle 0^\mathbb{N}, 0^\mathbb{N}, \ldots\rangle, n0^\mathbb{N}\rangle \in \dom(H)$.

If $\langle \langle 0^\mathbb{N}, 0^\mathbb{N}, \ldots\rangle, n0^\mathbb{N}\rangle \notin \dom(H)$, then $n0^\mathbb{N} \notin \left (\sum_{k+1}^\infty \llpo \right )(K(\langle 0^\mathbb{N}, 0^\mathbb{N}, \ldots\rangle))$ must hold. This in turn implies $K_2^n(\langle 0^\mathbb{N}, 0^\mathbb{N}, \ldots\rangle) \neq 0^\mathbb{N}$. Due to continuity of $K_2^n$, there must be some $m \in \mathbb{N}$ such that for all $p_i \in \Baire$ we have $K_2^n(\langle 0^mp_1, 0^mp_2, \ldots\rangle) \neq 0^\mathbb{N}$. Noting $\llpo_{k, 1}(\langle 0^mp_1, 0^mp_2, \ldots\rangle) = \llpo_{k, 1}(\langle p_1, p_2, \ldots\rangle)$, we see that $H$, $K$ also witness a reduction from $\llpo_{k, 1}$ to $\sum_{k+1}^\infty \llpo$ restricted to those $\langle q, \langle p_1, p_2, \ldots\rangle\rangle$ in its domain with $p_n \neq 0^\mathbb{N}$. It is easy to see that the latter is equivalent to $\llpo_{\infty, 1}$, hence we have a reduction from $\llpo_{k, 1}$ to $\llpo_{\infty, 1}$, but this contradicts \cite[Satz 18]{mylatzb}.

\item Define $h : \{n \mid n > 1\} \to \{i \mid 1 \leq i \leq k\}$ via $H(\langle \langle 0^\mathbb{N}, 0^\mathbb{N}, \ldots\rangle, n0^\mathbb{N}\rangle) = h(n)0^\mathbb{N}$.

By 1. together with the restrictions on the range of $H$, $h(n)$ is defined for $n > 1$, and has to take a value in $\{i \mid 1 \leq i \leq k\}$.

\item $h$ is injective.

     Assume we have $n \neq m$ with $h(n) = h(m)$. By continuity of $H$, there is some $l \in \mathbb{N}$ with $H(\langle \langle 0^lp_1, 0^lp_2, \ldots\rangle, n0^\mathbb{N}\rangle) = H(\langle \langle 0^lp_1, 0^lp_2, \ldots\rangle, m0^\mathbb{N}\rangle) = h(n)0^\mathbb{N}$ for all suitable $p_i$. This can only produce a correct answer to $\llpo_{k, 1}$, if $K_2^n(\langle 0^lp_1, 0^lp_2, \ldots\rangle) \neq 0^\mathbb{N} \neq K_2^m(\langle 0^lp_1, 0^lp_2, \ldots\rangle)$ for $p_{h(n)} \neq 0^\mathbb{N}$. But then $K$ only produces an element of $\dom(\sum_{k+1}^\infty \llpo)$ for such input, if $K_1(\langle 0^lp_1, 0^lp_2, \ldots\rangle) = 0^\mathbb{N}$ for $p_{h(n)} \neq 0^\mathbb{N}$. By continuity of $K_1$, then also $K_1(\langle 0^lp_1, \ldots, 0^lp_{h(n)-1}, 0^\mathbb{N}, 0^lp_{h(n)+1}, \ldots\rangle) = 0^\mathbb{N}$ follows. Noting again $\llpo_{k, 1}(\langle 0^{l'}p_1, 0^{l'}p_2, \ldots\rangle) = \llpo_{k, 1}(\langle p_1, p_2, \ldots\rangle)$, we obtain a reduction from $\llpo_{k, 1}$ to $\llpo_{\infty, 2}$, contradicting \cite[Satz 18]{mylatzb}.

\item By the pigeon hole principle, 2. and 3. are mutually exclusive. This contradicts the initial assumption.\qedhere
\end{enumerate}
\end{proof}

\begin{thm}
Neither $\mathfrak{C}_0$ nor $\mathfrak{C}$ is a Brouwer algebra.
\end{thm}

\begin{proof}
We shall prove that $\{R \mid \left ( \sum_{2}^\infty \llpo \right ) \leq_W^c R \coprod \llpo_{2, 1}\} := \mathcal{R}$ contains no minimal element.
\begin{enumerate}[(1)]
\item $\left ( \sum_{k}^\infty \llpo \right ) \in \mathcal{R}$ for all $k > 1$.

For $k = 2$ the claim is obvious. Now observe $( \sum_{k}^\infty \llpo ) \equiv_W \llpo_{k, 1} \coprod ( \sum_{k+1}^\infty \llpo )$, recall $\llpo_{k + 1, 1} <_W \llpo_{k, 1}$ and proceed by induction.

\item For any $R \in \mathcal{R}$, there is a $k > 1$ with $\left ( \sum_{k}^\infty \llpo \right ) \leq_W^c R$.

Let $\left ( \sum_{2}^\infty \llpo \right ) \leq_W^c R \coprod \llpo_{2, 1}$ be witnessed by continuous $H$, $K$. Then $A = \{p \in \dom(\sum_{2}^\infty \llpo) \mid K(p)(0) = 0\}$ and $B = \{p \in \dom(\sum_{2}^\infty \llpo) \mid K(p)(0) = 1\}$ are a relatively clopen disjoint cover of $\dom(\sum_{2}^\infty \llpo)$. One of the two sets must contain $\langle 0^\mathbb{N}, \langle 0^\mathbb{N}, 0^\mathbb{N}, \ldots, \rangle \rangle$ and hence also all $\langle 0^kp_0, \langle 0^kp_1, 0^kp_2, \ldots \rangle\rangle \in \dom(\sum_{2}^\infty \llpo)$ for some $k \in \mathbb{N}$. Thus, it follows that either $\left ( \sum_{k}^\infty \llpo \right ) \leq_W^c R$ or $\left ( \sum_{k}^\infty \llpo \right ) \leq_W^c \llpo_{2, 1}$. The latter assumption would imply $\llpo_{\infty, 2} \leq_W^c \llpo_{2, 1}$ and hence contradict \cite[Satz 18]{mylatzb}, leaving only the first case possible.

\item For any $k > 1$, we have $\left ( \sum_{k + 1}^\infty \llpo \right ) <_W^c \left ( \sum_{k}^\infty \llpo \right )$.

That $\left ( \sum_{k + 1}^\infty \llpo \right ) \leq_W^c \left ( \sum_{k}^\infty \llpo \right )$ holds is obvious, and a reduction in the other direction would contradict Lemma \ref{lemmallpo}.

\item By 1. and 2., any minimal $R \in \mathcal{R}$ would need to be equivalent to $\left ( \sum_{k}^\infty \llpo \right )$ for some $k \in \mathbb{N}$. But this would contradict 1. and 3., so there cannot be a minimal element in $\mathcal{R}$.\qedhere
\end{enumerate}
\end{proof}

\noindent As the next step, we demonstrate that none of the computable lattices, i.e. $\mathfrak{W}$, $p\mathfrak{W}$, $\mathfrak{W}^*$ and $p\widehat{\mathfrak{W}}$, is a Heyting algebra. Our proof makes use of the meet-reducibility of $\id_\Baire$ in $\mathfrak{W}$, so as contrast we present:

\begin{prop}
$\id_\Baire$ is meet-irreducible in $\mathfrak{C}$ (and hence in $\mathfrak{C}_0$, $\mathfrak{C}^*$ and $\widehat{\mathfrak{C}}$).
\label{propidirreduciblec}
\end{prop}

\begin{proof}
Assume $P, Q : \subseteq \Baire \mto \Baire$ such that $P \oplus Q$ has a continuous choice function $I$. Define $A = \{p \in \Baire \mid I(p)(0) = 0\}$ and $B = \{p \in \Baire \mid I(p)(0) = 1\}$. Then $A, B$ is a disjoint clopen cover of $\{\langle p, q\rangle \mid p \in \dom(P), q \in \dom(Q)\}$. If there is some $p_0 \in \dom(P)$ with $\langle p_0, q\rangle \in B$ for all $q \in \dom(Q)$, then $q \mapsto \textsc{Sh}(I(\langle p_0, q\rangle))$ is a continuous choice function for $Q$.

The negation of this assumption states that for all $p \in \dom(P)$ there is some $q_p$ with $\langle p, q_p\rangle \in A$. As $A$ is clopen, we can ensure that $q_p$ depends only on some finite prefix of $p$. Hence $p \mapsto q_p$ is continuous, and $p \mapsto \textsc{Sh}(I(\langle p, q_p\rangle))$ is a continuous choice function for $P$.

Being meet-irreducible in an upper cone is clearly sufficient for being meet-irreducible, thus the result for $\mathfrak{C}_0$ follows. Meet-irreducibility is a meet-semilattice property inherited by appropriate substructures, so the result translates to $\mathfrak{C}^*$ and $\widehat{\mathfrak{C}}$.
\end{proof}

\begin{thm}
Neither $\mathfrak{W}$ nor $p\mathfrak{W}$ is a Heyting algebra.
\label{notheyting1}
\end{thm}

\begin{proof}
Define $P : \{0^\mathbb{N}, 1^\mathbb{N}\} \to \{0^\mathbb{N}, J(0^\mathbb{N})\}$ via $P(0^\mathbb{N}) = 0^\mathbb{N}$ and $P(1^\mathbb{N}) = J(0^\mathbb{N})$. Then $\{R \mid P \oplus R \leq_W \id_\Baire\}$ has no maximal (pointed) element. To see this, for any $q \in \Baire$ define $Q_q : \{0^\mathbb{N}, J(0^\mathbb{N})\} \to \{0^\mathbb{N}, q\}$ via $Q_q(0^\mathbb{N}) = 0^\mathbb{N}$ and $Q_q(J(0^\mathbb{N})) = q$.

$P \oplus Q_q$ is computable (and pointed, hence equivalent to $\id_\Baire$): If either of the two arguments is $0^\mathbb{N}$, then the index of the respective argument together with $0^\mathbb{N}$ is a valid output and can easily be produced. If this is not the case, the input must be $\langle 1^\mathbb{N}, J(0^\mathbb{N})\rangle$. But from that, we can of course compute $0J(0^\mathbb{N})$, which constitutes a valid output.

So if $\overline{R}$ were a maximal element in $\{R \mid P \oplus R \leq_W \id_\Baire\}$, we would have $Q_q \leq_W \overline{R}$ for all $q \in \Baire$. All that a computable reduction to $\overline{R}$ can see from it are the values taken on inputs Turing reducible to $J(0^\mathbb{N})$, i.e.~it follows that $Q_q \leq_W \overline{R}_{|\{p \in \Baire \mid p \leq_T J(0^\mathbb{N})\}}$ for any $q \in \Baire$. But as $\{p \in \Baire \mid p \leq_T J(0^\mathbb{N})\}$ is countable, only countably many Turing degrees are below any $\overline{R}(p)$ with $p \leq_T J(0^\mathbb{N})$. This yields the desired contradiction.
\end{proof}

\begin{cor}
Neither $\mathfrak{W}^*$ nor $p\widehat{\mathfrak{W}}$ is a Heyting algebra.
\end{cor}

\begin{proof}
Using \cite[Proposition 4.9]{brattka2} the proof of Theorem \ref{notheyting1} can readily be adapted.
\end{proof}

\begin{cor}
$\id_\Baire$ is not meet-irreducible in $\mathfrak{W}$, $p\mathfrak{W}$, $\mathfrak{W}^*$ or $p\widehat{\mathfrak{W}}$.
\end{cor}

\begin{proof}
$P$ and $Q_{J(J(0^\mathbb{N}))}$ from the proof of Theorem \ref{notheyting1} form a counterexample.
\end{proof}

Not only do we find that the continuous versions of the lattices actually are Heyting algebras, but -- even more surprisingly -- our proof of this fact primarily makes use of recursion theoretic methods. While the interpretation of continuity as computability with respect to an arbitrary oracle \cite{ziegler2, weihrauchd} generally does allow recursion theory to be applied to continuous functions, to our knowledge this is the first application of recursion theoretic methods in the continuous case which does not constitute a relativization of the corresponding result for the computable case.

\begin{defi}
\label{defimpl}
For $P, Q : \subseteq \Baire \mto \Baire$ with $P \nleq_W^c Q$ and $P \neq \top$, we define $(P \rightarrow Q) : \subseteq \Baire \mto \Baire$ via:
\[\eqalign{\dom(P \rightarrow Q) = \{ijq \mid &\forall p \in \dom(P)
  \ \Phi_i(\langle p, q\rangle) \in \dom(Q) \wedge \cr&\forall r \in
  Q(\Phi_i(\langle p, q\rangle)) \ \Phi_j(\langle \langle p, q\rangle,
  r\rangle) \in 0P(p) \cup 1\Baire\}
}
\]
\[(P \rightarrow Q)(ijq) = \bigcup_{p \in \dom(P)} \bigcup_{r \in
    Q(\Phi_i(\langle p, q\rangle))} \{\textsc{Sh}(s) \mid s =
  \Phi_j(\langle \langle p, q\rangle, r\rangle) \wedge s \in
  1\Baire)\}
\]
For $P \leq_W^c Q$, we set $P \rightarrow Q = \top$, and understand $\top \rightarrow Q = Q$.
\end{defi}

To see that $P \rightarrow Q$ is actually well-defined, we have to verify $(P \rightarrow Q)(ijq) \neq \emptyset$ for $P \nleq_W^c Q$ and $ijq \in \dom(P\rightarrow Q)$. Assuming  $(P \rightarrow Q)(ijq) = \emptyset$ for such $ijq$, we see that $\Phi_i(\langle p, q\rangle) \in \dom(Q)$ for all $p \in \dom(P)$, and $\Phi_j(\langle \langle p, q\rangle, r\rangle) \in 0P(p)$ for all $r \in Q(\Phi_i(\langle p, q\rangle))$. But that means that relative to the oracle $q$ we have $P \leq_W Q$, hence we indeed find $P \leq_W^c Q$ in contradiction to our assumption.

\begin{thm}
\label{heyting2}
$\mathfrak{C}_0$ is a Heyting algebra, with $P \rightarrow Q$ being a
maximal element of \linebreak $\{R \in \mathfrak{C}_0 \mid P \oplus R
\leq_W^c Q\}$.
\end{thm}

\begin{proof}
In the case $P \leq_W^c Q$ or $P = \top$ the claim is clear. For $P \nleq_W^c Q$, we show $P \oplus (P \rightarrow Q) \leq_W Q$. For $p \in \dom(P)$ and $ijq \in \dom(P \rightarrow Q)$, we have $\Phi_{i}(\langle p, q\rangle) \in Q$. Given some $r \in Q(\Phi_{i}(\langle p, q\rangle))$, if $\Phi_j(\langle \langle p, q\rangle, r\rangle) = 0r'$, then $r' \in P(p)$ holds. If $\Phi_j(\langle \langle p, q\rangle, r\rangle) = 1r'$, then $r' \in (P \rightarrow Q)(ijq)$ holds.

Now assume $P \oplus R \leq_W^c Q$ witnessed by continuous functions $H, K$. Any continuous function is computable with respect to some oracle, hence we can assume $H(\langle \langle p_1, p_2\rangle, q\rangle) = H'(\langle \langle p_1, \langle p_2, O\rangle\rangle, q\rangle)$ and $K(\langle p_1, p_2\rangle) = K'(\langle p_1, \langle p_2, O\rangle\rangle)$ with computable functions $K', H'$ and some constant oracle $O \in \Baire$ by providing the oracle information in a suitable way.

From this, we can construct some $R'$ from $R$ with $P \oplus R' \leq_W Q$ being witnessed by computable $H', K'$, and $R(p) = R'(\langle p, O\rangle)$, hence $R \equiv_W^c R'$. Set $H' = \Phi_j$ and $K' = \Phi_i$. Then for any $q \in \dom(R')$ we have $ijq \in \dom(P \rightarrow Q)$.  Moreover, we find $(P \rightarrow Q)(ijq) \subseteq R'(q)$, which together implies $R' \leq_W (P \rightarrow Q)$. By transitivity, then also $R \leq_W^c (P \rightarrow Q)$ follows.
\end{proof}

The proof of the preceding theorem shows more than necessary. In fact, $\rightarrow$ has some properties of an implication even in $\mathfrak{W}$:

\begin{cor}
If $P\not\le^c_WQ$, then $P\oplus(P\to Q)\le_WQ$.
\end{cor}

\begin{cor}
If $P\oplus R\le_WQ$, then $R\le_W(P\to Q)$.
\end{cor}

\begin{cor}
$\mathfrak{C}$ is a Heyting algebra.
\end{cor}

\begin{proof}
Every upper cone in a Heyting algebra is a Heyting algebra itself.
\end{proof}

\begin{prop}
\label{proparrowtimes}
$(P \rightarrow Q_1) \times (P \rightarrow Q_2) \leq_W (P \rightarrow (Q_1 \times Q_2))$
\end{prop}

\proof
If $(P \rightarrow (Q_1 \times Q_2)) = \top$ or $P = \top$, then the claim is trivially true. Otherwise, we can conclude $P \nleq_W^c Q_1$, $P \nleq_W^c Q_2$, $P \nleq_W^c (Q_1 \times Q_2)$; so for all three occurrences of $\rightarrow$, the first case of Definition \ref{defimpl} is used then.

The reduction in the case $(P \rightarrow (Q_1 \times Q_2)) \neq \top
\neq P$ is witnessed by computable $H$, $K$, defined via $H(\langle r,
\langle p_1, p_2\rangle\rangle) = \langle p_1, p_2\rangle$ and
$K(\langle i_1j_1q_1, i_2j_2q_2\rangle) = ij\langle q_1, q_2\rangle$,
where $i$ is an index such that 
\[\Phi_i(\langle p, \langle q_1, q_2\rangle\rangle) = \langle
\Phi_{i_1}(\langle p, q_1\rangle), \Phi_{i_2}(\langle p,
q_2\rangle)\rangle
\]
  and $j$ is an index such that 
\[\eqalign{&\Phi_j(\langle \langle p, \langle q_1, q_2\rangle\rangle, \langle
r_1, r_2\rangle\rangle)\cr = 
&\begin{cases}1\langle \textsc{Sh}(\Phi_{j_1}(\langle
\langle p, q_1\rangle,r_1\rangle)), \textsc{Sh}(\Phi_{j_2}(\langle
\langle p, q_2\rangle, r_2\rangle))\rangle
 &\mbox{\ \ if\ }
  \Phi_{j_k}(\langle \langle p, q_k\rangle, r_k\rangle)\in
  1\mathbb{N}^\mathbb{N}\mbox{\ for\ }k=1,2\\
  \Phi_{j_1}(\langle \langle p, q_1\rangle,r_1\rangle)
 &\mbox{\ \ if\ }\Phi_{j_1}(\langle \langle p, q_1\rangle, r_1\rangle)(0)= 0\\
  \Phi_{j_2}(\langle \langle p, q_2\rangle, r_2\rangle)
 &\mbox{\ \ otherwise\rlap{\hbox to 133 pt{\hfill\qEd}}}\\
\end{cases}
}
\]

\begin{prop}
\label{proparrowdom}
Suppose $P \nleq_W^c Q$ and $P \neq \top$. Then $(P \rightarrow Q)$ is pointed iff $\dom(Q) \leq_M \dom(P)$.
\end{prop}

\begin{proof}
Let $ijq \in \dom(P \rightarrow Q)$ be computable. Then the computable
map $p \mapsto \Phi_i(\langle p, q\rangle)$ witnesses $\dom(Q) \leq_M
\dom(P)$.  Conversely, let computable $H$ witness $\dom(Q) \leq_M \dom(P)$, let $i_0$, $j_0$ be such that $\Phi_{i_0}(\langle p, q\rangle) = H(p)$ and $\Phi_{j_0}(r) = 1^\mathbb{N}$. Then $i_0j_0\Baire \subseteq \dom(P \rightarrow Q)$.
\end{proof}

\begin{cor}
$(P \rightarrow Q^*)^* \equiv_W (P \rightarrow Q^*)$
\end{cor}

\begin{proof}
If $P \leq_W^c Q^*$, then the claim evaluates to the trivially true $\top \equiv_W \top^*$. If $P = \top$, it becomes $(Q^*)^* \equiv_W Q^*$. In the following, assume $P \nleq_W^c Q^*$ and $P \neq \top$. As $Q^*$ is always pointed, by Proposition \ref{proparrowdom} the same holds for $(P \rightarrow Q^*)$. Together with $(P \rightarrow Q^*) \times (P \rightarrow Q^*) \leq_W (P \rightarrow Q^*)$ from Proposition \ref{proparrowtimes}, we see that $(P \rightarrow Q^*)$ is a fixed point of $^*$.
\end{proof}

\begin{cor}
$\mathfrak{C}^*$ is a Heyting algebra.
\end{cor}

Each of the three Heyting algebras we have identified gives rise to some superintuitionistic logic as its theory. As a starting point to determine those logics, we consider Jankov logic: By $\textsc{Jan}$ we denote the smallest superintuitionistic logic containing $\neg p \vee \neg \neg p$. Besides its simple definition, this logic is of interest as it arises as the theory of the dual of the Medvedev lattice $\mathfrak{M}^{op}$ \cite{jankov, medvedev, sorbi2}\footnote{In consulting these references, be wary that the theory of a lattice can refer to its theory as a Brouwer algebra, too, rather than its theory as a Heyting algebra as defined in the present paper. Such statements are mutually translatable by moving to the dual lattice.}.

An important property a superintuitionistic logic might exhibit is the disjunction property \cite{chagrov2}, which states that $p \vee q$ is true, if and only if $p$ or $q$ is true. In terms of Heyting algebras, this amounts to $\top$ being join-irreducible. In these cases, we can characterize those Heyting algebras validating the weak law of the excluded middle:

\begin{prop}
Let $(\mathfrak{L}, \wedge, \vee, \rightarrow)$ be a Heyting algebra with largest element $\top$ and smallest element $\bot$, such that $\top$ is join-irreducible. Then $\textsc{Jan} \subseteq \textsc{Th}(\mathfrak{L})$, if and only if $\bot$ is meet-irreducible.
\end{prop}

\begin{proof}
$\textsc{Jan} \subseteq \textsc{Th}(\mathfrak{L})$ amounts to $(\mathbf{a} \rightarrow \bot) \vee ((\mathbf{a} \rightarrow \bot) \rightarrow \bot) = \top$ for all $\mathbf{a} \in \mathfrak{L}$. As $\top$ is join-irreducible, this is equivalent to $(\mathbf{a} \rightarrow \bot) = \top$ or $((\mathbf{a} \rightarrow \bot) \rightarrow \bot) = \top$. By definition of $\rightarrow$, the former is equivalent to $\mathbf{a} \wedge \top = \mathbf{a} = \bot$. The latter is equivalent to $(\mathbf{a} \rightarrow \bot) = \bot$, which in turn is equivalent to $\mathbf{a} \wedge \mathbf{b} = \bot$ implies $\mathbf{b} = \bot$.

Clearly, the first case only holds for $\mathbf{a} = \bot$. The second case, formulated for $\mathbf{a} \neq \bot$, is equivalent to the meet-irreducibility of $\bot$.
\end{proof}

Due to the nature of the special functions representing $\bot$ and $\top$ in all our lattices, we see that they are join-irreducible and meet-irreducible in all of our lattices in which they occur. Together with Proposition \ref{propidirreduciblec} and Corollary \ref{propidirreduciblec} showing meet-irreducibility of $\id_\Baire$ in $\mathfrak{C}^*$ we get:

\begin{cor}
$\textsc{Jan} \subseteq \textsc{Th}(\mathfrak{C}_0)$, $\textsc{Jan} \subseteq \textsc{Th}(\mathfrak{C})$ and $\textsc{Jan} \subseteq \textsc{Th}(\mathfrak{C}^*)$
\end{cor}

In any of our Heyting algebras, we find $P \vee \neg P = P \neq \top$ for $P \neq \bot, \top$, so their theories are proper subsets of classical propositional logic; and neither of our lattices is a Boolean lattice. Better upper bounds could be obtained by embedding suitable Heyting algebras into $\mathfrak{C}_0$, $\mathfrak{C}$ and $\mathfrak{C}^*$.

If $\mathfrak{C}^*$ or $\widehat{\mathfrak{C}}$ should be a Brouwer algebra, or $\widehat{\mathfrak{C}}$ be a Heyting algebra, then the considerations above do equally apply to $\textsc{Th}((\mathfrak{C}^*)^{op})$, $\textsc{Th}(\widehat{\mathfrak{C}}^{op})$ and $\textsc{Th}(\widehat{\mathfrak{C}})$.

\section{Embedding the Medvedev degrees}
Embeddings of the Medvedev degrees into the Weihrauch degrees were first studied in \cite[Section 5]{brattka2}, using a definition very similar to the following:
\begin{defi}
Given some $\mathcal{A} \subseteq \Baire$, define $c_{\mathcal{A}} : \subseteq \Baire \mto \Baire$ via $\dom(c_\mathcal{A}) = \{0^\mathbb{N}\}$ and $c_\mathcal{A}(0^\mathbb{N}) = \mathcal{A}$.
\end{defi}

As shown in \cite{brattka2}, $c : \mathfrak{M} \hookrightarrow \mathfrak{W}$ is a meet-semilattice embedding, and $c : \mathfrak{M} \hookrightarrow p\widehat{\mathfrak{W}}$ is a lattice embedding (hence $c : \mathfrak{M} \hookrightarrow \mathfrak{W}^*$ is also a lattice embedding). First, we shall complement these results by a corresponding negative one.

\begin{prop}
\label{propcajoinirreducible}
$c_\mathcal{A}$ is join-irreducible for all $\mathcal{A} \subseteq \Baire$.
\end{prop}

\begin{proof}
Any reduction $c_\mathcal{A} \leq_W P \coprod Q$ witnessed by $H$, $K$ makes use of only the single point $K(0^\mathbb{N})$ in the domain of $P \coprod Q$. But then only one of $P$ and $Q$ is ever evaluated, hence, either $c_\mathcal{A} \leq_W P$ or $c_\mathcal{A} \leq_W Q$ follows.
\end{proof}

\begin{cor}
$c : \mathfrak{M} \hookrightarrow \mathfrak{W}$ is not a lattice embedding.
\end{cor}

The embedding $c$ even preserves the structure of $\mathfrak{M}$ as a Brouwer algebra, as far as its codomain permits. This is formalized as the following:

\begin{prop}
Let $\mathfrak{L}$ be a sublattice of $\mathfrak{W}^*$ with $c(\mathfrak{M}) \subseteq \mathfrak{L}$, such that $\mathfrak{L}$ is a Brouwer algebra. Then $c: \mathfrak{M} \hookrightarrow \mathfrak{L}$ is a Brouwer embedding.
\end{prop}

\begin{proof}
We show the even stronger result that for any $\mathcal{A}, \mathcal{B} \subseteq \Baire$ the set $\{R \in \mathfrak{W} \mid c_{\mathcal{A}} \leq_W c_\mathcal{B} \times R\}$ has a minimal element of the form $c_{\mathcal{C}}$. This in turn already follows from the claim that $c_{\mathcal{A}}\leq_W c_\mathcal{B} \times R$ implies the existence of some $\mathcal{C} \subseteq \Baire$ with $c_\mathcal{C} \leq_W R$ and $c_{\mathcal{A}} \leq_W c_\mathcal{B} \times c_\mathcal{C}$, together with $\mathfrak{M}$ being a Brouwer algebra and $c : \mathfrak{M} \to \mathfrak{W}^*$ being a lattice embedding.

To see the latter claim, assume that $c_{\mathcal{A}}\leq_W c_\mathcal{B} \times R$ is witnessed by $H$, $K$. We have $K(0^\mathbb{N}) = \langle 0^\mathbb{N}, p\rangle$ for some computable $p \in \Baire$. Let $\mathcal{C} = R(p)$. Then $c_\mathcal{C} \leq_W R$ is obvious, and $c_{\mathcal{A}} \leq_W c_\mathcal{B} \times c_\mathcal{C}$ is witnessed by $H$, $K'$ where $K'(0^\mathbb{N}) = \langle 0^\mathbb{N}, 0^\mathbb{N}\rangle$.
\end{proof}

There is another natural embedding of the Medvedev degrees into the Weihrauch degrees, this time however the ordering is reversed. We shall consider the straight-forward definition originally suggested by \textsc{Brattka}:
\begin{defi}
Given some $\mathcal{A} \subseteq \Baire$, define $d_{\mathcal{A}} : \subseteq \Baire \to \Baire$ via $\dom(d_\mathcal{A}) = \mathcal{A}$ and $d_\mathcal{A}(x) = 1^\mathbb{N}$ for $x \in \mathcal{A}$.
\end{defi}

\begin{lem}
$d : \mathfrak{M}^{op} \hookrightarrow \mathfrak{W}$ is a lattice embedding.
\end{lem}

\begin{proof}
Recall that the lattice operations in $\mathfrak{M}$ are $+$ and $\times$ defined via $\mathcal{A} + \mathcal{B} = 0\mathcal{A} \cup 1\mathcal{B}$ and $\mathcal{A} \times \mathcal{B} = \{\langle p, q\rangle \mid p \in \mathcal{A} \wedge q \in \mathcal{B}\}$.
\begin{enumerate}[(1)]
\item $\mathcal{A} \leq_M \mathcal{B}$ iff $d_\mathcal{B} \leq_W d_\mathcal{A}$

Let computable $K$ witness $\mathcal{A} \leq_M \mathcal{B}$. Then $d_{\Baire}$ and $K$ witness $d_\mathcal{B} \leq_W d_\mathcal{A}$. Conversely, let $H$, $K$ witness $d_\mathcal{B} \leq_W d_\mathcal{A}$. Then $K$ must witness $\mathcal{A} \leq_M \mathcal{B}$.

\item $d_{\mathcal{A} \times \mathcal{B}} \equiv_W d_\mathcal{A} \oplus d_\mathcal{B}$

Both directions are witnessed by $H = d_\Baire$, $K = \id_\Baire$.

\item $d_{\mathcal{A} + \mathcal{B}} \equiv_W d_\mathcal{A} \coprod d_\mathcal{B}$

Both directions are witnessed by $H = d_\Baire$, $K = \id_\Baire$.\qedhere
\end{enumerate}
\end{proof}

\noindent The following observation shows that the image $d(\mathfrak{M}^{op})$ contains exactly the computable Weihrauch degrees:
\begin{obs}
$P : \subseteq \Baire \mto \Baire$ has a computable choice function, iff $P \equiv_W d_{\dom(P)}$.
\end{obs}

Expanding on this, we see that the computable Weihrauch degrees are isomorphic to the dual of the Medvedev degrees. Taking into consideration that the computable Weihrauch degrees are the lower cone of $\id_\Baire$, we cannot hope to expand the Heyting algebra structure of $\mathfrak{M}^{op}$ any further from this starting point.

\section{Summary}
Finally, we summarize our results which lattices also have the structure of a Heyting or Brouwer algebra. Note the three cases remaining open. It seems that additional techniques will be necessary to provide the corresponding answers.
$$\begin{array}{l|cccccccc}
 & \mathfrak{W} & p\mathfrak{W} & \mathfrak{W}^* & p\widehat{\mathfrak{W}} & \mathfrak{C}_0 & \mathfrak{C} & \mathfrak{C}^* & \widehat{\mathfrak{C}} \\
 \hline
 \textnormal{Brouwerian?} & \textsc{No} & \textsc{No} & \textsc{No} & \textsc{No} & \textsc{No} & \textsc{No} & ? & ? \\
 \textnormal{Heyting?} & \textsc{No} & \textsc{No} & \textsc{No} & \textsc{No} & \textsc{Yes} & \textsc{Yes} & \textsc{Yes} & ?
\end{array} $$

\section*{Acknowledgement}
We would like to thank Vasco Brattka for introducing us to the questions discussed in this paper.
\end{document}